\documentclass[prb,aps,a4paper,superscriptaddress,twocolumn,showpacs]{revtex4}
\usepackage{graphicx,latexsym}
\usepackage{dcolumn}
\usepackage{amsmath,amssymb,epsf,bm}

\begin{document}

\title{Spin Injection in Quantum Wells with Spatially Dependent Rashba
Interaction}
\author{Arne Brataas}
\affiliation{Centre for Advanced Study at the Norwegian Academy of Science and Letters,
Drammensveien 78, NO-0271 Oslo, Norway}
\affiliation{Department of Physics, Norwegian University of Science and Technology,
NO-7491 Trondheim, Norway}
\author{A. G. Mal'shukov}
\affiliation{Centre for Advanced Study at the Norwegian Academy of Science and Letters,
Drammensveien 78, NO-0271 Oslo, Norway}
\affiliation{Institute of Spectroscopy, Russian Academy of Science, 142190 Troitsk,
Moscow Region, Russia}
\author{Yaroslav Tserkovnyak}
\affiliation{Centre for Advanced Study at the Norwegian Academy of Science and Letters,
Drammensveien 78, NO-0271 Oslo, Norway}
\affiliation{Department of Physics and Astronomy, University of California, Los Angeles,
California 90095, USA}

\begin{abstract}
We consider Rashba spin-orbit effects on spin transport driven by an
electric field in semiconductor quantum wells. We derive spin
diffusion equations that are valid when the mean free path and the
Rashba spin-orbit interaction vary on length scales larger than the
mean free path in the weak spin-orbit coupling limit. From
these general diffusion equations, we derive boundary conditions between regions of different
spin-orbit couplings. We show that spin injection is feasible when
the electric field is perpendicular to the boundary between two
regions. When the electric field is parallel to the boundary, spin
injection only occurs when the mean free path changes within the
boundary, in agreement with the recent work by Tserkovnyak
\textit{et al.} [cond-mat/0610190].
\end{abstract}

\maketitle


\section{Introduction}

Spintronics envisions electronic devices with spin injection,
detection, and manipulation. The spin-orbit interaction couples the
electron spin to its orbital motion. It was early realized that it
can be utilized to manipulate the electron spin in e.g.\ the
Datta-Das spin transistor.\cite{Datta:apl90} More recently, the
possibility of spin injection via the spin-orbit interaction in
non-magnetic systems was
suggested.\cite{Hirsch:prl99,Murakimi:science03} The spin-orbit
interaction gives rise to the spin Hall effect where a longitudinal
electric field induces a transverse spin current. In semiconductors,
the spin Hall effect has generated a large interest and is predicted
to occur in a wide variety of electron and hole doped 2D and 3D
systems. \cite{Sinova:prl04} A resulting accumulation of a
normal to 2DEG spin polarization near sample boundaries has recently
been measured.\cite{Kato:science04} Also, injection of the spin Hall
polarization into a region of the 2DEG where the driving electric
field is absent has been observed in Ref.\ \onlinecite{SihAwschalom
_injection}.

The spin Hall effect is conventionally separated into an intrinsic and an
extrinsic effect. The intrinsic spin Hall effect is caused by the spin-orbit
splitting of energy bands in zinc-blende semiconductors. In contrast, the
extrinsic spin Hall effect arises due to spin-dependent scattering off
impurities. This paper considers the intrinsic spin Hall effect in the
diffusive transport regime for electrons confined in a quantum well subject
to the Rashba spin-orbit interaction. We will focus on spin diffusion in systems where the Rashba coupling
constant $\alpha$ varies in space. In particular, we address a problem of
injection of the spin polarization from regions with a large spin-orbit coupling $\alpha$ to
regions where this constant is small or zero.

Recent research has established how spins diffuse, precess and
couple to the charge transport in electron doped III-V semiconductor
quantum wells. The charge and spin flows are described by
spin-charge coupled diffusion equations.
\cite{Burkov:prb04,Mischenko:prl04} However, the diffusion
equations, valid in the bulk of the systems, must be supplemented by
appropriate boundary conditions. Calculations to this end were
carried out in
Refs.\ %
\onlinecite{Adagideli:prl05,Malshukov:prl05,Galitski:prb06,Bleibaum:prb06.1,Bleibaum:prb06.2,Rashba:physica06}
and have given rise to discussion in recent literature.
The derivation of these boundary conditions is more subtle than it
might first seem since the calculations must be carried out to the
second order in the spin-orbit interaction strength and include
the effects of the electric field.

The correct boundary condition when the electric field is parallel
to the boundary and the spin-orbit coupling varies in space
was derived in Ref. \onlinecite{Tserkovnyak}. An argument in terms
of conventional Hall physics that extends the results to systems
where also the mobility varies was also given: A
spatially dependent spin-orbit coupling varying on length scales
smaller than the spin-precession length corresponds to an
effective magnetic field of opposite magnitude for two spin
directions.\cite{Tserkovnyak,Tserkovnyak:cm0611086} The effective magnetic field induces spin dependent
Hall voltage across the boundary.\cite{Tserkovnyak} This additional voltage,
in its turn, gives rise to the spin density variation in the
transition layer, resulting in a finite spin-density jump, when the transition layer thickness
is reduced to zero.

This paper aims to address  spin diffusion and the boundary
conditions in more detail and generality for the Rashba spin-orbit
interaction model. Our aim is to clarify some conceptual issues and
not to fit experimental data because typical experimental
systems are rather complex with competing effects and
material parameters that are not well known. The Rashba spin-orbit
interaction in semiconductor quantum wells depends on quantum well
confinement, which can be manipulated by gates, or doping profile,
\cite{Nitta97} but usually only at rather long scales. The length
scales for variations in system properties such as the spin-orbit
interaction and the mean free path are typically much larger than
the mean free path. Therefore, we will derive general diffusion
equations that are valid when the system parameters vary on length
scales longer than the mean free path. These diffusion equations can
then be solved for smooth boundaries between regions of different
spin-orbit coupling strengths and mean free path. From these
solutions, we will find boundary conditions that are not limited to
a specific relative direction between the electric field and the
interface. When the electric field is parallel to the interface our
derived boundary conditions agree with the result by Tserkovnyak
\textit{et al.}\cite{Tserkovnyak} and the analog with conventional
Hall physics is rigorously justified beyond the simplest scenario
considered microscopically in Ref.~\onlinecite{Tserkovnyak}. We also
present results for the boundary conditions when the electric field
is perpendicular to the interface.

Our paper is organized in the following way. In the next section \ref{boundary}, 
we summarize our main results for the resulting boundary
conditions between regions with different spin-orbit couplings and mean free
path. We discuss the implications of these boundary conditions on
spin injection and show that for a parallel electric field spin injection is
only possible when the mean free path together with $\alpha$ vary within the
boundary layer, while for the perpendicular field spin injection takes place
even when the mean free path is uniform in the system. Using the Keldysh formalism, we derive the general
diffusion equation in section \ref{diffusion} that we employ to compute the boundary conditions presented
in section \ref{boundary}. Our conclusions are in
section \ref{conclusions}.

\section{Boundary conditions and spin injection}

\label{boundary}

In the absence of disorder, the electrons in a homogeneous quantum well are described
by the Hamiltonian%
\begin{equation}
H=\frac{\hbar ^{2}\mathbf{k}^{2}}{2m^{\ast }} +\bm{\sigma }\cdot \mathbf{%
h}_{\mathbf{k}},  \label{H}
\end{equation}%
where $m^{\ast }$ is the electron effective mass,
$\mathbf{k}=\left( k_{x},k_{y}\right) $ is the wave vector,
$\bm{\sigma}$ is a vector of Pauli matrices, and
$\mathbf{h}_{\mathbf{k}}$ is the effective spin-orbit field. Let
us assume that the major contribution to the spin-orbit
interaction (SOI) is given by the Rashba interaction:
\begin{equation}
h_{\mathbf{k}}^{x}=\alpha k_{y}\,\,\,,\,\,\,h_{\mathbf{k}}^{y}=-\alpha
k_{x}\,,  \label{hRashba}
\end{equation}%
where $k_{i}$ is the 2D electron wave vector. We also assume that
the spin-orbit coupling constant $\alpha (\mathbf{r})$ is
spatially modulated. In this case, the spin orbit term in Eq.\ (\ref{H})
has to be modified to preserve the Hermitian form of the
Hamiltonian, namely, this term is rewritten as 1/2 of the
anticommutator of the momentum operator and the spatially
dependent $\alpha (\mathbf{r})$.

Since the spatial variations of $\alpha$ are assumed to be provided by an
appropriate quantum well modulation or impurity doping profile, any
consistent model must in general also take into account spatial variations of the mean
free scattering time $\tau (\mathbf{r})$. The electron gas is under the action
of the electric field $\mathbf{E}(\mathbf{r})$ and our goal is to find the spin
density induced in the 2DEG by $\mathbf{E}$. The electric field may vary in regions where the mobility changes
in order to maintain a constant charge current density.  We will derive the general
diffusion equations corresponding to the Hamiltonian (\ref{H}) in the next
section (\ref{diffusion}).

\begin{figure}[hh]
  \includegraphics[width=\linewidth]{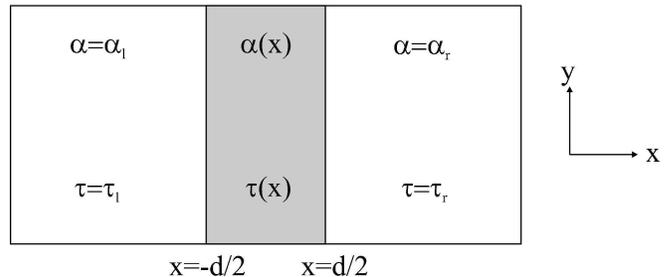}
  \caption{A 2DEG with a spatial dependent Rashba interaction and scattering time.
The Rashba interaction $\alpha(x)$ and scattering time $\tau(x)$ vary in the transition region between the left
region where $\alpha=\alpha_l$ and $\tau=\tau_l$ and the right region where $\alpha=\alpha_r$ and $\tau=\tau_r$.
}
  \label{fig:2DEG}
\end{figure}

Let us first discuss some of our main findings of the general diffusion
equations when applied to the problem of spin injection
from a region with a large spin-orbit coupling to a region with a
smaller spin-orbit coupling.
As an example, we consider a left ($x<-d/2$) and right region ($x>d/2$)
with different spin-orbit
couplings and mean free paths connected via a smooth transition layer, as shown in Figure \ref{fig:2DEG}.
We assume that the transition layer thickness $d$ between the regions is larger
than the mean free path $l$, but smaller than the spin precession length $%
\lambda=\hbar^2/m^*\alpha$, where $m^*$ is the electron effective mass, 
$l\ll d\ll \lambda$. This requires a sufficiently weak spin-orbit
interaction or a low mobility sample. We consider the
general case where the electric field is in the two dimensional
plane, $\mathbf{E}=E^{x}\mathbf{x}+E^{y}\mathbf{y}$.

The diffusion equations for the spin-components $S^{x}$, $S^{y}$, and $S^{z}$
of the spin accumulation in the bulk of each region ($x<-d/2$ or $x>d/2$)
valid when $1/\tau,h_{k_{F}}\ll E_{F}$ are \cite{Burkov:prb04,Mischenko:prl04,Malshukov:prl05}
\begin{eqnarray}
D\nabla ^{2}S^{x}+4Dm^{\ast }\alpha \nabla S^{z} &=&\Gamma _{\parallel
}\left( S^{x}-S_{b}^{x}\right) \,,  \nonumber\\
D\nabla ^{2}S^{y} &=&\Gamma _{\parallel }(S^{y}-S_{b}^{y}) \,,  \nonumber\\
D\nabla ^{2}S^{z}-4Dm^{\ast }\alpha \nabla S^{x} &=&\Gamma _{\perp }S^{z}\,,
\label{Sdiff}
\end{eqnarray}%
where $D=v_{F}^{2}\tau /2$ is the diffusion constant in terms of the
Fermi velocity $v_{F}$ and the elastic scattering time $\tau $ and
$\Gamma _{\perp }=2\Gamma _{\parallel }=8D\left( m^{\ast }\right)
^{2}\alpha ^{2}$ are the D'yakonov-Perel spin relaxation rates. It
is well known that in systems with Rashba interaction, a uniform
electric field induces a nonequilibrium bulk spin polarization
parallel to the 2DEG,\cite{Edelstein} denoted in
Eqs.~(\ref{Sdiff}) as $S_{b}^{x}$ and $S_{b}^{y}$. Within the linear response theory, the spin polarization
components are
\begin{eqnarray*}
S_{b}^{x} &=&-\alpha eE^{y}\tau N_{F}, \\
S_{b}^{y} &=&\alpha eE^{x}\tau N_{F},
\end{eqnarray*}
where $N_{F}=m^{\ast}/(\hbar^2 2\pi)$ is the density of states at the
Fermi energy. The spin current is not a conserved quantity in
spin-orbit coupled systems, but, with the conventional definition
within the semiclassical theory,\cite{Tang:prb05} it reads in
diffusive systems for spin flow, e.g., along the $x$
direction:\cite{Malshukov:prl05}
\begin{eqnarray*}
j^{x} &=&D\left[ \nabla _{x}S^{x}+2m^{\ast }\alpha S^{z}\right] , \label{jx} \\
j^{y} &=&D\nabla _{x}S^{y}, \label{jy} \\
j^{z} &=&D\left[ \nabla _{x}S^{z}-2m^{\ast }\alpha \left(
S^{x}-S_{b}^{x}\right) \right]\,. \label{jz}
\end{eqnarray*}
We generalize below the diffusion equations (\ref{Sdiff})
to systems with spatially varying parameters $\alpha$ and
$\tau$.

By solving these generalized diffusion equations in the boundary transition
layer ($-d/2<x<d/2$), that is thinner than the spin-precession length $%
\lambda$, we find in Sec.~III the boundary conditions
\begin{eqnarray}
\left( S^{x}-S_{b}^{x}\right) _{d/2}-\left( S^{x}-S_{b}^{x}\right) _{-d/2}
&=&- \! \int_{-d/2}^{d/2} \! \! \! \! \! \! \! \! dx S_{b}^{x}\frac{1}{\tau }%
\frac{d\tau }{dx}\,,  \label{Sx_bound} \\
S^{y}_{d/2}-S^{y}_{-d/2} &=& 0 ,,  \label{Sy_bound} \\
S_{d/2}^{z}-S_{-d/2}^{z} &=&0\,.  \label{Sz_bound}
\end{eqnarray}
The boundary condition for the $S^x$ spin density component can be understood in terms of a conventional
Hall effect for two spin directions in the transition layer, see the discussion in Ref.\
 \onlinecite{Tserkovnyak}. We outline in the next
section \ref{diffusion} how the boundary condition for the $S^y$ component can be understood as a result of electron drift
in an effective spin dependent potential.

The boundary condition for the spatial derivative of the spin
density is particularly simple. It can be written as continuity of spin current in its conventional definition:
\begin{equation}
\left( \mathbf{j}_{s}\right) _{d/2}-\left( \mathbf{j}_{s}\right) _{-d/2}=0
\label{Cur_bound}
\end{equation}%
Naively, the continuity of the spin current across the boundary is
not surprising because the boundary conditions have been obtained
assuming that the boundary layer is thinner than the spin precession
length. That means that effects violating spin current conservation
only occur on much larger length scales. Consequently, there is spin
current conservation through the boundary. However, the continuity
of the spin current across the boundary cannot be rigorously 
established without doing an actual calculation. There
could in general be a violation of the spin current conservation of
the order $\alpha^{2}$, but we find explicitly in the next section
\ref{diffusion} that also such corrections are absent.

Let us now discuss the implications of the diffusion equations (\ref{Sdiff})
and the boundary conditions (\ref{Sx_bound}), (\ref{Sy_bound}), (\ref{Sz_bound}), and (\ref{Cur_bound}) for
spin injection from a region with e.g. a larger spin-orbit coupling
constant to a region with a smaller spin-orbit coupling constant. We consider two limits for the alignment between the electric field and the boundary:
1) The electric field is parallel to the boundary, $\mathbf{E}=E^{y}\mathbf{y%
}$, $E^{x}=0$ and 2) the electric field is perpendicular to the boundary, $%
\mathbf{E}=E^{x}\mathbf{x}$, $E^{y}=0$. Since we consider the linear response
regime, spin injection for general orientations between the
electric field and the boundary can be found as a linear combination of 1)
and 2).

The former case 1) of a parallel electric field $\mathbf{E}=E^{y}\mathbf{y}$
corresponds to the case studied by Tserkovnyak~\textit{et al.}\cite{Tserkovnyak} For this geometry, the
bulk nonequilibrium spin accumulation along the $y$ direction vanishes, $%
S_{b}^{y}=0$, and the boundary condition (\ref{Sy_bound}) implies that
\begin{equation*}
S^{y}=0
\end{equation*}%
throughout the system. The only nonvanishing bulk component of the spin
distribution is along $x$. Deep inside the bulk of the left (l) and right (r)
regions:
\begin{eqnarray*}
S_{b_{l}}^{x} &=&-\alpha _{l}eE^{y}\tau _{l}N_{F}, \\
S_{b_{r}}^{x} &=&-\alpha _{r}eE^{y}\tau _{r}N_{F}.
\end{eqnarray*}%

If the electron mobility is the same in each region, $\tau_{l}=\tau_{r}$, and constant
throughout the boundary, then the spin accumulation $S^{x}$ will
change rapidly (i.e., on the scale of the spatial variations of
$\alpha$, $d$) from $S_{b_{l}}^{x}=-\alpha
_{l}eE^{y}\tau _{l}N_{F}$ at $x=-d/2$ to $S_{b_{r}}^{x}=-\alpha _{r}eE^{y}\tau
_{r}N_{F}$ at $x=d/2$.
For the slowly varying $S^{x}$ in bulk, this looks as a
discontinuity in the spin accumulation at the boundary. When the
mobility differs in the two regions, the magnitude of
discontinuity in the spin accumulation across
the boundary is modified according to Eq.~(\ref{Sx_bound}). This result differs from
Ref.~\onlinecite{Adagideli:prl05}, where a continuous spin density across the
boundary was assumed. Solving the diffusion equations (\ref{Sdiff}),
the solution in the left region is ($x<-d/2$)
\cite{Burkov:prb04}%
\begin{eqnarray*}
S^{x} &=&S_{b_{l}}^{x}+\sum_{n=1,2}S_{i_{l},n}^{x}\exp q_{l,n}\left(
x+d/2\right)\,, \\
S^{z} &=&\sum_{n=1,2}S_{i_{r},n}^{z}\exp q_{l,n}\left( x+d/2\right)
\end{eqnarray*}%
and in the right region ($x>d/2$)
\begin{eqnarray*}
S^{x} &=&S_{b_{r}}^{x}+\sum_{n=1,2}S_{i_{r},n}^{x}\exp q_{r,n}\left(
-x+d/2\right)\,, \\
S^{z} &=&\sum_{n=1,2}S_{i_{r},n}^{z}\exp q_{r,n}\left( -x+d/2\right) ,
\end{eqnarray*}%
where $q_{n}$  is the solution with positive real part of $q_{n}=(1/\lambda)\sqrt{2(-1\pm i\sqrt{7})}$.
Continuity of the spin current, Eq.~(\ref{Cur_bound}), combined with the boundary conditions for $S^{x}$ and $S^{z}$, Eqs.~(\ref{Sx_bound}) and (\ref{Sz_bound}), imply
\begin{eqnarray*}
S_{i_{l},n}^{x} &=&\varkappa _{l,n}^{x}\Delta S^{x}\,, \\
S_{i_{r},n}^{x} &=&\varkappa _{r,n}^{x}\Delta S^{x}\,, \\
S_{i_{l},n}^{z} &=&\varkappa _{l,n}^{z}\Delta S^{x}\,, \\
S_{i_{r},n}^{z} &=&\varkappa _{r,n}^{z}\Delta S^{x}\,,
\end{eqnarray*}%
where%
\begin{equation*}
\Delta S^{x}=-\!\int_{-d/2}^{d/2}\!\!\!\!\!\!\!\!dxS_{b}^{x}(x)\frac{1}{\tau }%
\frac{d\tau }{dx}
\end{equation*}
and $\varkappa _{l}^{x}$, $\varkappa _{r}^{x}$, $\varkappa^z_l$ and $\varkappa^{z}_r$ are
lengthy algebraic dimensionless coefficients that depend on $q_{l,n}$,  $%
q_{r,n}$ and the diffusion coefficients $D_{l}$ and $D_{r}$. From
these boundary conditions, it should be clear that spin injection,
which we define as a spin accumulation that differs from the bulk
spin density ${\bf S}_b$, i.e., $S_i \ne 0$, is only
feasible when the mobility
 varies within the boundary
layer, since otherwise $\Delta S^{x}$ vanishes. If the mobility is
constant throughout the system,  there is no spin injection
from the left region to the right region or vice versa. In
general, $\Delta S^{x}$ depends on the specifics of the variation
of the mobility and the spin-orbit coupling in the boundary layer.

Next, let us study the second scenario 2) where the electric field is
perpendicular to the boundary, $\mathbf{E}=E^{x}\mathbf{x}$, $E^{y}=0$. In
this case, we will find that spin injection is feasible even when the
mobility is uniform throughout the system. In the bulk of the left (right)
region, we have $S^{y}=S_{b_{l}}^{y}=\alpha _{l}eE^{x}_l\tau
_{l}N_{F}$ ($S^{y}=\alpha _{r}eE^{x}_r\tau_{r}N_{F}$), while the other spin components vanish. 
Since the electric current is continous across the interface, the electric fields on both sides of the boundary
must differ in order to adjust to the different mobilities so that $E^x_l \tau_l = E^x_r \tau_r$. 
The boundary conditions (\ref{Sx_bound}), (\ref{Sz_bound}), and (\ref{Cur_bound}) then imply that
\begin{equation*}
S^{x}=0\,\,\, {\rm and}\,\,\, S^{z}=0
\end{equation*}
throughout the system. In contrast to the previous scenario 1),
where the in-plane spin component $S^{x}$ exhibits a discontinuity
across the transition layer when the mobility is uniform, $S^{y}$
is continous here. Solving the diffusion equations (\ref{Sdiff}),
we find in the left region
($x<-d/2$)
\begin{equation*}
S^{y}=S_{b_{l}}^{y}+S_{i_{l}}^{y}\exp \lambda _{l}^{-1}\left( x+d/2\right) ,
\end{equation*}%
and in the right region ($x>d/2$)
\begin{equation*}
S^{y}=S_{b_{r}}^{y}+S_{i_{r}}^{y}\exp \lambda _{r}^{-1}\left( -x+d/2\right) ,
\end{equation*}%
Continuity of the spin current and the boundary condtions for $S^{y}$ imply
\begin{eqnarray*}
S_{i_{l}}^{y} &=&\varkappa _{l}^{y}\left[ -\left(
S_{b_{r}}^{y}-S_{b_{l}}^{y}\right) \right]  \\
S_{i_{r}}^{y} &=&\varkappa _{r}^{y}\left[ -\left(
S_{b_{r}}^{y}-S_{b_{l}}^{y}\right) \right]
\end{eqnarray*}%
where
\begin{eqnarray*}
\varkappa _{l}^{y} &=&-\frac{D_{r}\lambda _{r}^{-1}}{D_{l}\lambda
_{l}^{-1}+D_{r}\lambda _{r}^{-1}}, \\
\varkappa _{r}^{y} &=&\frac{D_{l}\lambda _{l}^{-1}}{D_{l}\lambda
_{l}^{-1}+D_{r}\lambda _{r}^{-1}} \, .
\end{eqnarray*}%
In the limit of weak spin-flip relaxation in the left region,
$\lambda _{l}\gg \lambda _{r}D_{l}/D_{r}$, we find $\varkappa
_{l}^{y}=-1$, $\varkappa _{r}^{y}=0$ and $\varkappa ^{z}=0$. In
contrast to scenario 1), spin injection is feasible even when the
mobility is constant throughout the system. This means that spin injection is feasible independent 
on the spatial variation of mobilities.

Let us now compare the spin-injection efficiency in scenario 1)
(electric field parallel to the interface)
and 2) (electric field perpendicular to the interface). It
is clear that geometry 2) is more effective than geometry 1) when the
mobility is constant throughout the system. For the
simplest case, where both the mean free path and the spin-orbit
coupling linearly change within the boundary layer:
\begin{eqnarray*}
\alpha(x) &
= & \frac{\alpha
_{r}+\alpha _{l}}{2}+\frac{x}{d} \left( \alpha _{r}-\alpha _{l}\right) \, , \\
\tau(x) & = & \frac{ \tau _{r}+\tau _{l}}{2}+\frac{x}{d} \left( \tau _{r}-\tau
_{l}\right) \, ,
\end{eqnarray*}
we find
\begin{equation*}
\Delta S^{x}=\frac{\alpha _{r}+\alpha _{l}}{2}N_{F}eE^{y}\left( \tau _{r}-\tau _{l}\right)
\end{equation*}
and
\begin{equation}
-\left(
S_{b_{r}}^{y}-S_{b_{l}}^{y}\right) =-\left( \alpha
_{r}-\alpha _{l}\right) N_{F}\frac{eE^{x}_r \tau _{r}+ eE^{x}_l \tau
_{l}}{2} \, .
\end{equation}
For this simplest system of a
linear variation of the spin-orbit coupling and mobility and assuming that
one uses the same electric fields in the two scenarios, then 2) is more
effective than 1) provided%
\begin{equation*}
\left\vert \frac{\alpha _{l}-\alpha _{r}}{\alpha _{l}+\alpha _{r}}%
\right\vert >\left\vert \frac{\tau _{l}-\tau _{r}}{\tau _{l}+\tau _{r}}%
\right\vert ,
\end{equation*}%
e.g., when the change in the spin-orbit coupling constant is larger than the
change in mobility between the regions. Here, we are crudely characterizing the
spin-injection efficiency by the jump in spin density with respect to the bulk levels.

\section{Derivation of Diffusion Equation \label{diffusion}}

We will in this section derive general diffusion equations that are valid
when the mean free path and spin-orbit coupling vary on length scales longer
than the mean free path.
It is well known that in a homogeneous system an electric field induces a
spin polarization parallel to the 2DEG.\cite{Edelstein} In an inhomogeneous
system, an interplay of this effect and the spin Hall effect can give rise to
a more complicated spin distribution. A convenient tool
to study such a system is the spin diffusion equation. We start from the
Keldysh formalism and obtain the Boltzmann equation for the spin
distribution function. In terms of the Boltzmann function, the spin density
is defined as $\mathbf{S}=\sum_{\mathbf{k}}\bm{g}_{\mathbf{k}}$, and the
nonequilibrium charge density as $2\sum_{\mathbf{k}}g_{\mathbf{k}}^{\text{ch}%
}$. The linearized equation for the nonequilibrium part of the Boltzmann function can be written
in the form (see e.g. Refs.~\onlinecite{Tang:prb05,Tserkovnyak})
\begin{eqnarray}
&&\mathbf{v}\mathbf{\nabla }g_{\mathbf{k}}^{i}-\alpha (\hat{z}\times \mathbf{%
\nabla })_{i}g_{\mathbf{k}}^{\text{ch}}+2(\bm{g}_{\mathbf{k}}\times \bm{h}_{%
\mathbf{k}})_{i}  \notag  \label{Boltzmann} \\
&&-\frac{\partial g_{\mathbf{k}}^{\text{ch}}}{\partial \mathbf{k}}\mathbf{%
\nabla }h_{\mathbf{k}}^{i}+e\mathbf{E}\frac{\partial g_{\mathbf{k}}^{i(0)}}{%
\partial \mathbf{k}}=\frac{1}{\tau }\left( S_{E}^{i}-g_{\mathbf{k}%
}^{i}\right) \,,
\end{eqnarray}
where $\mathbf{v}$=$\mathbf{k}/m^*$, $i=x,y,z$, $\mathbf{S}_{E}=\delta (E_{\mathbf{k}}-E_{F})\mathbf{S}%
/N_{F}$, $\bm{g}_{\mathbf{k}}^{(0)}=-\bm{h}_{\mathbf{k}}\delta (E_{%
\mathbf{k}}-E_{F})$ is the equilibrium Boltzmann function for spin, and $%
\hat{z}$ is the unit vector perpendicular to 2DEG. The first two
terms on the left-hand side  of Eq.\ (\ref{Boltzmann})
describe variations of the Boltzmann function due to
particle motion with spin-dependent velocity, the 3rd term is the
spin precession in the Rashba field, the 4th term is the
spin-dependent acceleration in the spatially dependent Rashba field,
and 5th term is associated with the driving electric field.
The right hand side of Eq.\ (\ref{Boltzmann}) represents collisions
in the self-consistent Born approximation caused by weak and
isotropic scalar disorder.

In the leading approximation, the charge component of the Boltzmann function is determined
by the drift of electrons in the external field:
\begin{equation}
g_{\mathbf{k}}^{\text{ch}}=e\tau \frac{\mathbf{k}\mathbf{E}}{m^{\ast }}%
\delta (E_{\mathbf{k}}-E_{F})  \label{gcharge}
\end{equation}
which can be spatially dependent through $\tau \equiv \tau (\mathbf{r})$. Due to this dependence,
the second term on the left hand side of Eq. (\ref{Boltzmann}) becomes important in the following
calculation of the spin polarization. We note that this takes place only for the electric field parallel
to the interface. For a perpendicular field, as we have mentioned above, the combination $\tau E^x$ is 
continous across the boundary.

The next step is to derive from (\ref{Boltzmann}) the diffusion equation for
the spin polarization. There are two ways to do that, depending on how fast $%
\tau (\mathbf{r})$ and $\alpha (\mathbf{r})$ vary in space. If the
corresponding length scale is shorter than the electron mean free
path $l$, the diffusion equation can be applied only within
regions where variations of $\tau (\mathbf{r})$ and $\alpha
(\mathbf{r})$ are slow, while the regions of their fast modulation
have to be described by the original kinetic equation
(\ref{Boltzmann}). For example, one can consider a thin transition
layer between two regions characterized by nonequal uniform
values of $\tau (\mathbf{r})$ and $\alpha (\mathbf{r})$. In this case, Eq.~(\ref%
{Boltzmann}) can be used to derive boundary conditions for diffusion
equations on both sides of the interface. If variations of the parameters
take place on length scales larger than $l$, the general diffusion equation valid
in the entire 2D system can be straightforwardly derived from Eq.~(\ref%
{Boltzmann}). We will follow the second procedure assuming slow enough
variations of $\tau (\mathbf{r})$ and $\alpha (\mathbf{r})$, throughout the system.
We will see in the following that the characteristic spatial scale of the spin density variations
dictated by the diffusion equation is the spin precession length $\lambda
=\hbar^2/m^{\ast }\alpha$. Hence, this length must be much larger than $l$,
which means also that $h_{\mathbf{k}}\ll 1/\tau $. Then, following Ref.~\onlinecite{Tang:prb05},
the diffusion equation can be directly
obtained from Eq.~(\ref{Boltzmann}). To this end, let us denote
\begin{equation}
Q^{i}=\frac{S_{E}^{i}}{\tau }+\alpha (\hat{z}\times \mathbf{\nabla })_{i}g_{%
\mathbf{k}}^{\text{ch}}+\frac{\partial g_{\mathbf{k}}^{\text{ch}}}{\partial
\mathbf{k}}\mathbf{\nabla }h_{\mathbf{k}}^{i}-e\mathbf{E}\frac{\partial g_{%
\mathbf{k}}^{i(0)}}{\partial \mathbf{k}}\,.  \label{Q}
\end{equation}%
The function $g_{\mathbf{k}}^{i}$ can now be found by applying the operator
\begin{equation}
\hat{\Lambda}^{-1}\approx \tau \left[ 1-(\bm{v}\mathbf{\nabla })\tau +(\bm{v}%
\mathbf{\nabla })\tau (\bm{v}\mathbf{\nabla })\tau \right] .  \label{Lambda}
\end{equation}%
which is the inverse to $\hat{\Lambda}=\bm{v}\mathbf{\nabla }+\tau ^{-1}$ expanded to the
second order with respect to $\tau \bm{v}\mathbf{\nabla }$, where $\bm{v}$
is the particle velocity. In this way, we obtain
\begin{equation}
\bm{g}_{\mathbf{k}}=\hat{\Lambda}^{-1}\left[ \bm{Q}-2h_{\mathbf{k}}(\bm{g}_{\mathbf{k}%
}\times \bm{n}_{\mathbf{k}})\right]  \,, \label{g}
\end{equation}
where $\bm{n}_{\mathbf{k}}=\bm{h}_{\mathbf{k}}/h_{\mathbf{k}}$,
while to order $\alpha ^{2}$ the vector product in (\ref{g}) can be
expressed as
\begin{equation}
\bm{g}_{\mathbf{k}}\times \bm{n}_{\mathbf{k}}=\hat{\Lambda}^{-1}(\bm{Q}%
\times \bm{n}_{\mathbf{k}})+2\tau ^{2}h_{\mathbf{k}}\bm{Q}_{\perp
}\,, \label{perp}
\end{equation}
where $\bm{Q}_{\perp
}=\bm{Q}-\bm{n}_{\mathbf{k}}(\bm{n}_{\mathbf{k}}\bm{Q})$. In the $h_{%
\mathbf{k}}^{2}$ terms we disregarded gradient corrections. Furthermore, the $%
\hat{\Lambda}^{-1}$ operator in Eqs.~(\ref{perp}) and (\ref{g}) can
be expanded according to Eq. (\ref{Lambda}). Below, we will keep the
terms up to order $\nabla^2$ and $\alpha^3$ in expansions in
$\nabla$ and $\alpha$, as well as up to $\alpha^2\nabla$
 in cross products. Substituting
Eq.~(\ref{perp}) and Eq.~(\ref{Q}) into Eq.~(\ref{g}) and taking a
sum over $\mathbf{k}$, we arrive at a closed diffusion equation for
the spin density. For simplicity, we consider the case when $\alpha
$ and $\tau $ depend only on the $x$ coordinate, while the direction
of the electric field is arbitrary. This is sufficient to discuss,
e.g., a boundary between two regions with different Rashba
couplings. Finally, the system of diffusion equations takes the form
\begin{widetext}
\begin{eqnarray}
\nabla_x D \nabla_x (S^x-S^x_b) +\nabla_x D S^x_b
\frac{\tau^{\prime}}{\tau}+2 D m^{*}\alpha \nabla_x S^z + 2
\nabla_x D
m^{*}\alpha S^z - \Gamma_{\parallel} (S^x-S^x_b)&=&0 \label{diffx} \\
\nabla_x D \nabla_x S^y- \Gamma_{\parallel} (S^y-S^y_b)&=&0
\label{diffy} \\
\nabla_x D \nabla_x S^z - 2D m^{*}\alpha S^x_b
\frac{\tau^{\prime}}{\tau}-2 D m^{*}\alpha \nabla_x(S^x-S^x_b) - 2
\nabla_x D m^{*}\alpha (S^x-S^x_b) - \Gamma_{\perp} S^z&=&0 \label{diffz}\,,
\end{eqnarray}
\end{widetext}
where $\tau ^{\prime }=d\tau/dx $.

Let us consider the limiting case when the mobility through the
scattering time $\tau$ is homogenous, $\tau'=0$. It can then be seen
that the solution of Eqs.~(\ref{diffx}) and (\ref{diffz})
 is $S^z=0$, and $S^x=S^x_b$. Since $S^x_b$ depends on $x$
through $\alpha(x)$, this spin density component simply follows the
spatial dependence of the Rashba coupling constant, as it was
discussed in the previous section. The behavior of $S^y$ is
different. If, for example, $\alpha$ varies from some finite value
to zero across the transition layer, Eq.~(\ref{diffy}) shows that
the electric field perpendicular to the interface, $E_x$, drives
injection of $S^y$ into a region of \textit{e.g.} a vanishing Rashba
spin-orbit interaction, corresponding to scenario 2) analyzed in
detail in section \ref{boundary}.

Let us consider a thin transition layer where $\alpha (x)$ and $\tau (x)$
vary between their constant values on the left and on the right from this
layer. In this case,  the spin polarization is expected to change fast within
the layer, while it varies much more slower outside the layer, if its width $%
d$ is much less than the spin-precession length $\lambda$. The magnitude of the spin density variation
across the layer can be found directly from the diffusion equations (\ref{diffx})-(\ref{diffz}). Indeed, in the
leading approximation with respect to $d/\lambda$, when $x$ is within the
layer, we retain in the diffusion equations only the leading terms proportional to $%
\nabla _{x}^{2}$. After integration of the first of the equations from a
point $x_{1}$ just to the left from the interface to some point $x$,
it becomes
\begin{equation}
\nabla _{x}(S^{x}-S_{b}^{x})+S_{b}^{x}\frac{\tau ^{\prime }}{\tau }-[\nabla
_{x}(S^{x}-S_{b}^{x})+S_{b}^{x}\frac{\tau ^{\prime }}{\tau }]_{x=x_{1}}.
\label{bc1}
\end{equation}%
The expression in square brackets can be set to 0 because $\nabla
_{x}S^{x}|_{x=x_{1}}\sim (S^{x})/\lambda,\nabla _{x}S_{b}^{x}|_{x=x_{1}}=0$,
and $S_{b}^{x}\tau ^{\prime }/\tau |_{x=x_{1}}=0$. Hence, integration of Eq.~(%
\ref{bc1}) from $x_{1}$ to $x_{2}$ yields
\begin{equation}
(S^{x}-S_{b}^{x})|_{x=x_{2}}-(S^{x}-S_{b}^{x})|_{x=x_{1}}=-%
\int_{1}^{2}dxS_{b}^{x}\frac{\tau ^{\prime }}{\tau } \,. \label{bc2}
\end{equation}
Similarly, it is easy to show that the $z$ spin component slowly
varies across the boundary, while the $y$ spin component changes according to
\begin{equation}
S^{y}|_{x=x_{2}}-S^{y}|_{x=x_{1}}= 0. \label{bc3}
\end{equation}
Equation (\ref{bc2}) coincides with the result obtained in Ref.\ \onlinecite{Tserkovnyak}.
It shows that when $\tau $ is $x$ dependent, $%
S^{x} $ does not follow exactly the spatial profile of $S_{b}^{x}$. Due to
this mismatch, according to Eq.~(\ref{diffz}), the $S^{z}$
component of the spin density is not zero on both sides of the transition
layer and spin injection is feasible.

The rest of the boundary conditions can be obtained from
Eqs.~(\ref{diffx}), (\ref{diffy}), and (\ref{diffz}) by
integrating these equations across the layer and taking into account that $%
S^{z}$ and the sum of the first two terms in (\ref{bc1}) vary slowly
within the layer. Therefore, the term proportional to $\nabla
_{x}S^{z}$ can be disregarded when integrating
Eq.~(\ref{diffx}) and also can be disregarded the sum of
the second and third terms of the last line. They give rise only to
corrections $\sim \alpha^2 d/\lambda$. Also, taking into account that outside the layer
$\tau ^{\prime }=\alpha ^{\prime }=0$, we obtain
\begin{align}
\left[ \nabla _{x}S^{x}+2m^{\ast }\alpha S^{z}\right] _{x_{2}} &=\left[
\nabla _{x}S^{x}+2m^{\ast }\alpha S^{z}\right] _{x_{1}}\,,  \notag \\
\left[ \nabla _{x}S^{z}-2m^{\ast }\alpha (S^{x}-S_{b}^{x})\right] _{x_{2}}&=\left[ \nabla _{x}S^{z} - 2m^{\ast }\alpha(S^{x}-S_{b}^{x})\right]_{x_{1}}\,,
\notag \\
\nabla _{x}S^{y}|_{x_{2}} &=\nabla _{x}S^{y}|_{x_{1}}\,.
\end{align}
Hence, we arrive at the boundary conditions corresponding to spin current conservation
across the interface in its conventional definition (up to the order $\alpha^{2}$),
although in general such a current is not a conserved quantity.

\section{Conclusions \label{conclusions}}

We have derived general diffusion equations for spin transport driven by
electric fields in a quantum well subject to Rashba spin-orbit interaction. The
diffusion equations are valid for spatially dependent mean free paths and
spin-orbit interaction couplings. From these general diffusion equations, we
can determine the boundary conditions for the spin-density and spin-current
components in regions with different mean free paths and spin-orbit couplings
via boundaries that are thicker than the mean free path, but thinner than
the spin-precession length. We find that spin injection is possible when the
electric field is perpendicular to the interface, and also when the electric
field is parallel to the interface. In the latter case, spin injection is
possible provided the boundary layer has a spatially dependent mobility.

We thank B.\ I.\ Halperin for an important comment. This work was supported in part by the 
Research Council of Norway through Grant Nos. 158518/143, 158547/431, and 167498/V30 and the 
Russian \ RFBR through grant No. 060216699.

\end{document}